\documentclass[10pt,conference]{IEEEtran}
\usepackage{graphicx}
\usepackage{tcolorbox}
\usepackage{multirow}
\usepackage{makecell}
\usepackage{threeparttable}
\usepackage{array}
\usepackage{listings}
\usepackage{tabularx}
\usepackage{xcolor}
\usepackage{colortbl}
\usepackage{fancyvrb} 
\usepackage{fontawesome}
\usepackage{enumitem}
\usepackage{array}
\usepackage{booktabs} 
\usepackage{pifont}
\usepackage[font=small]{caption} 
\usepackage{titlesec}
\usepackage{balance}
\usepackage{xspace}
\usepackage{algorithmic}
\usepackage{amsmath,amsfonts}

\definecolor{c1}{HTML}{F6D4CF}
\definecolor{tabHeader}{HTML}{EAF0F6}
\definecolor{tabSubHeader}{HTML}{F6F8FA}
\definecolor{boundRow}{HTML}{EAF6EF}
\definecolor{tabRule}{HTML}{4B5563}
\newcolumntype{Y}{>{\centering\arraybackslash}X}

\newcommand{\appname}{{BOUND}\xspace}

\newif\ifSimpleMode\SimpleModefalse

\IEEEoverridecommandlockouts
\begin{document}

\title{Mitigating Package Hallucinations in Large Language Models via Model Editing}

\author{
\IEEEauthorblockN{
Shuhan Liu\IEEEauthorrefmark{1},
Yukai Zhao\IEEEauthorrefmark{1},
Xing Hu\IEEEauthorrefmark{1}\textsuperscript{\ddag},
Kui Liu\IEEEauthorrefmark{2},
Xiaohu Yang\IEEEauthorrefmark{1},
and Xin Xia\IEEEauthorrefmark{1}
}
\IEEEauthorblockA{
\IEEEauthorrefmark{1}The State Key Laboratory of Blockchain and Data Security, Zhejiang University, Hangzhou, China\\
\{liushuhan, yukaizhao2000, xinghu, yangxh\}@zju.edu.cn, xin.xia@acm.org
}
\IEEEauthorblockA{
\IEEEauthorrefmark{2}Huawei, Hangzhou, China
brucekuiliu@gmail.com
}
\thanks{\textsuperscript{\ddag} Corresponding Author}
}

\maketitle

\begin{abstract}
Large language models (LLMs) have demonstrated strong capabilities in software engineering tasks, such as code generation, library recommendation, and dependency configuration. 
However, recent studies show that LLMs may suffer from package hallucination, where they generate non-existent or invalid package names.
These hallucinations can be exploited in software supply chain attacks, as attackers may register malicious packages under hallucinated names. 
Therefore, mitigating package hallucination is important for improving the reliability and security of LLM-assisted software development.
In this paper, we introduce \appname, a lightweight localized model editing framework for mitigating package hallucinations in LLMs. \appname formulates package hallucination mitigation as a package-validity boundary editing problem, where the boundary refers to the model's ability to distinguish valid packages from hallucinated package names under a given task context.
It first locates modules related to package hallucination through a risk-aware localization strategy, and then edits these modules with lightweight LoRA adapters using a boundary-aware objective that reinforces valid packages, suppresses hallucinated packages, and preserves locality behavior.
Experimental results show that \appname effectively reduces package hallucinations while preserving valid package recommendations. In the package recommendation task, \appname reduces package-level hallucination rate (Package-HR) by 79.9\% on edit prompts and by 65.4\% on unseen prompts. The learned package-validity boundary further generalizes to other package-related tasks, reducing Package-HR by 12.8\% in code generation and by 34.0\% in \texttt{pip install} recommendation.
These results show that \appname refines the package-validity boundary of LLMs and improves the reliability of package-related outputs.
\end{abstract}

\section{Introduction}
In recent years, large language models (LLMs) have developed rapidly and demonstrated strong capabilities across various software engineering tasks~\cite{wang2023review,jimenez2024swe,roziere2023code,sun2025source}. Developers increasingly rely on LLMs not only to generate code, but also to recommend libraries, configure dependencies, and prepare development environments~\cite{alhanahnah2024depsrag,latendresse2025robust}. In these workflows, package names generated by LLMs are often treated as actionable recommendations: they may be copied into import statements, installed through package managers, or used by automated coding agents. Consequently, the reliability of package-related outputs has become an important requirement for the practical adoption of LLMs in software development~\cite{pearce2025asleep,perry2023users}.

However, recent studies have shown that LLMs may suffer from hallucinations, generating outputs that appear plausible but are factually incorrect~\cite{zhang2025siren,gao2025current,zhang2025llm}.
One particularly concerning type is \emph{package hallucination}, where models generate non-existent or invalid software package names~\cite{spracklen2025we,krishna2025importing,zhao2025hfuzzer}. Unlike many conventional code generation errors, package hallucinations can directly affect dependency management and software supply chain security~\cite{ohm2020backstabber,ladisa2023sok,neupane2023beyond}. Developers and automated agents may inadvertently install hallucinated packages generated by LLMs, creating opportunities for attackers to register malicious packages under the same names~\cite{spracklen2025we}. Such risks are closely related to package confusion attacks, a well-known threat to software supply chains~\cite{lanyado2024diving,kaplan2021survey}. As LLM-based coding assistants and autonomous agents are increasingly integrated into development workflows, mitigating package hallucination is essential for building trustworthy software engineering environments.

Existing studies have proposed various strategies to mitigate package hallucinations, ranging from external mitigation techniques to internal model adaptation approaches~\cite{spracklen2025we}. External mitigation methods primarily operate outside the model. One line of work augments generation with external knowledge sources (e.g., PyPI~\cite{pypi}) to ground model outputs in reliable package information~\cite{lewis2020retrieval,jain2025mitigating}. Another line of work encourages the model to refine its own responses through self-correction~\cite{ji2023towards,madaan2023self}. Although these methods can reduce hallucinations, they do not directly modify the model’s internal behavior and introduce additional inference overhead or pipeline complexity~\cite{huang2025survey}.
Internal adaptation methods (e.g., supervised fine-tuning) provide a more direct alternative by modifying the model itself~\cite{spracklen2025we,tonmoy2024comprehensive}. However, fine-tuning updates a large number of parameters and may overfit to specific output formats, making it computationally expensive and potentially degrading the model’s general capabilities. These limitations motivate the need for a lightweight approach that can directly mitigate package hallucinations in LLMs without extensive retraining or architectural modifications.

Recently, model editing, also known as knowledge editing, has been proposed to update specific knowledge or behaviors in LLMs without full retraining~\cite{li2025model,wang2024detoxifying,zhang2024comprehensive,liu2025creme}. 
This provides a promising direction for mitigating package hallucination.
However, recent studies have shown that existing editing methods may still struggle to correct hallucinated outputs~\cite{huang2025can}.
Existing model editing methods primarily focus on factual knowledge updates, such as modifying subject-object associations or correcting individual facts~\cite{dai2022knowledge,li2024pmet,meng2022rome,wu2023depn}. 
These methods often rely on clear subject tokens or target phrases to locate and edit the corresponding knowledge. Package hallucination presents a different challenge. Package-related prompts usually contain complex task descriptions rather than explicit factual subjects, and the desired behavior is not to generate a fixed target output. Instead, the model should suppress non-existent package names while preserving valid packages. Furthermore, valid and hallucinated packages may appear together in the same response. This suggests that package hallucination is better viewed as a package-validity boundary problem than a factual rewriting problem. We define the package-validity boundary as the model's ability to distinguish valid packages (i.e., packages that exist in the package ecosystem) from hallucinated packages under a given task context, which is represented by the dashed separation line in Figure~\ref{fig:motivation}.

In this paper, we introduce \textbf{\appname}, a lightweight localized model editing framework for mitigating package hallucination in LLMs. Unlike existing mitigation methods that rely on external package retrieval or full-model fine-tuning, \appname aims to refine the package-validity boundary within the model itself. 
Specifically, \appname first employs a risk-aware localization strategy to identify modules whose parameters are most sensitive to the model’s preference for hallucinated over valid packages, which serves as a signal of the package-validity boundary. \appname then injects lightweight LoRA adapters into the localized modules and optimizes them with a boundary-aware objective that reinforces valid packages, suppresses hallucinated packages, and preserves the original model behavior on locality prompts.

To evaluate the effectiveness of \appname, we conduct experiments on a package hallucination dataset derived from package-related prompts~\cite{spracklen2025we}. We evaluate three representative open-source LLMs: \texttt{DeepSeekCoder}, \texttt{Qwen3}, and \texttt{Llama-3.1}. To provide a comprehensive evaluation, we compare \appname with five baselines, including two package hallucination mitigation methods (i.e., Full-FT~\cite{spracklen2025we} and Self-Refinement~\cite{spracklen2025we}) and three model editing methods (i.e., ROME~\cite{meng2022rome}, MEMIT~\cite{meng2023memit}, and DINM~\cite{wang2024detoxifying}).
Experimental results show that \ding{182} \appname effectively mitigates package hallucinations in the package recommendation task, reducing package-level hallucination rate (Package-HR) by 79.9\% on the edit set, and by 65.4\% on unseen prompts. \ding{183} The learned package-validity boundary generalizes to other package-related tasks, reducing Package-HR by 12.8\% in code generation and by 34.0\% in \texttt{pip install} recommendation. \ding{184} Compared with existing baselines, \appname achieves a better balance between hallucination reduction and valid package preservation, while several baselines reduce hallucinations by suppressing package generation. 
\ding{185} Ablation results show that both module localization and the three editing objectives are essential to effective editing.  
\ding{186} \appname is lightweight in practice, requiring only 731 KB of LoRA parameters on average, approximately 20,000$\times$ smaller than full fine-tuning checkpoints.

\noindent\textbf{Contributions:} The main contributions of this paper can be summarized as follows:
\begin{itemize}[leftmargin=*]
    \item We formulate package hallucination as a \emph{package-validity boundary} editing problem, aiming to refine the model's boundary between valid and hallucinated packages.
    \item We introduce \textbf{\appname}, a lightweight localized model editing framework that identifies modules related to package hallucination and applies boundary-aware LoRA editing to reduce hallucinated packages while preserving valid package recommendations.
    \item We conduct experiments on three open-source LLMs and show that \appname effectively reduces package hallucination, generalizes to other package-related tasks, and requires only lightweight editing artifacts.
\end{itemize}

\section{Background}
\label{sec:background}
\subsection{Motivating Example}
\begin{figure}[h]
  \centering  \includegraphics[width=\linewidth]{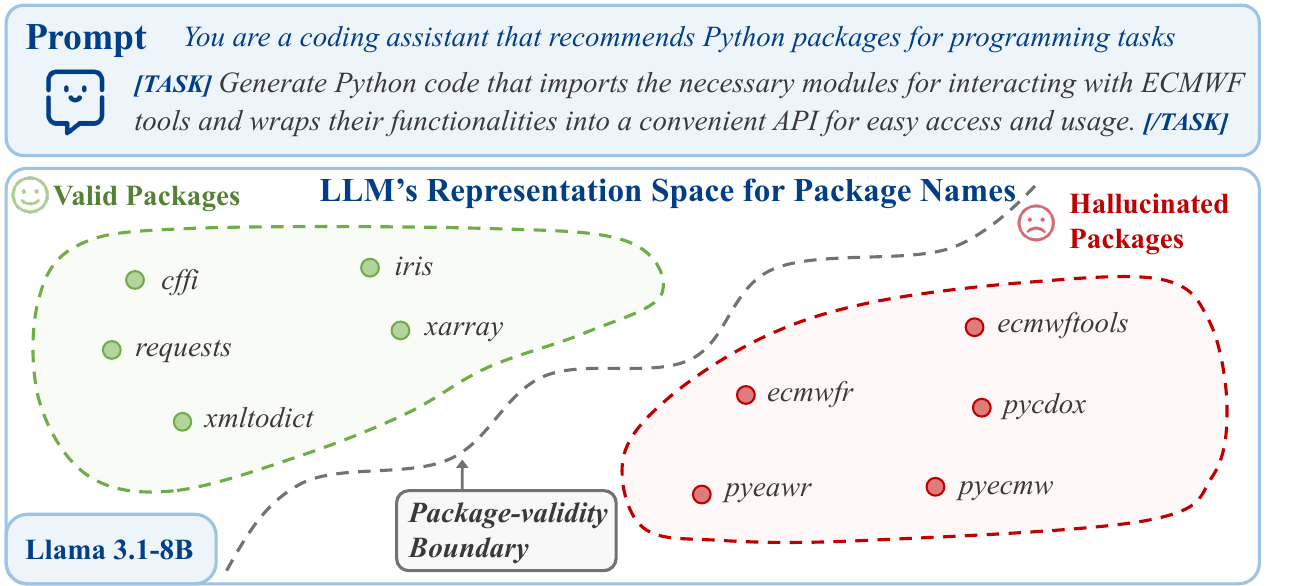}
  \caption{Motivating example of package hallucination as a package-validity boundary problem.}
    \label{fig:motivation}
\vspace{-0.1cm}
\end{figure}

Figure~\ref{fig:motivation} shows a representative package hallucination example from Llama-3.1-8B. 
Given a prompt asking for Python packages to interact with ECMWF tools and wrap their functionalities into a convenient API, the model generates both valid packages, such as \texttt{cffi} and \texttt{iris}, and non-existent package names, such as \texttt{ecmwfr} and \texttt{pyeawr}. 
Because these hallucinated names appear plausible under the task context, developers or automated coding agents may mistakenly treat them as valid dependencies, leading to installation failures or software supply chain risks if attackers register malicious packages under the same names.

This example shows that package hallucination is not simply caused by a lack of package knowledge. 
The model can identify some valid dependencies, but it still fails to distinguish them from plausible non-existent alternatives. 
We view package hallucination as a package-validity boundary error, where the model's distinction between valid and hallucinated packages is unreliable under a given task context. 
Consequently, mitigation should refine this boundary rather than rewrite a specific package recommendation into a fixed target.

\subsection{Task Definition}
We formulate the task of package hallucination mitigation through model editing in the context of package recommendation. 
Let $M_{\theta}$ denote a pretrained LLM with parameters $\theta$. Given a package recommendation prompt $x$, the model generates an answer $a \sim M_{\theta}(x)$, from which we extract a set of package names $P(a)$. Based on a package-validity checking procedure, the extracted packages are partitioned into valid packages $G(a)$ and hallucinated packages $H(a)$:
\begin{equation}
    P(a) = G(a) \cup H(a), \quad G(a) \cap H(a) = \emptyset .
\end{equation}

A package hallucination occurs when $H(a)\neq\emptyset$, meaning that the answer contains at least one hallucinated package. 
For the $i$-th case for model editing, let $a_i \sim M_{\theta}(x_i)$ be the original model response. 
We define the edit case as
\begin{equation}
    e_i = (x_i, G_i, H_i),
    \quad
    G_i = G(a_i),
    \quad
    H_i = H(a_i).
\end{equation}
where $G_i$ denotes valid package recommendations to be preserved or encouraged, and $H_i$ denotes hallucinated package recommendations to suppress.

The goal is to learn an edit $\Delta$ such that the edited model $M_{\theta+\Delta}$ reduces the likelihood of generating hallucinated packages while preserving valid package recommendations:
\begin{equation}
    M_{\theta+\Delta}(x_i) \rightarrow G_i,\quad
    M_{\theta+\Delta}(x_i) \not\rightarrow H_i.
\end{equation}

\section{Approach}
\label{sec:approach}
\subsection{Overview}
\begin{figure*}[t]
\centering
\includegraphics[width=\textwidth]{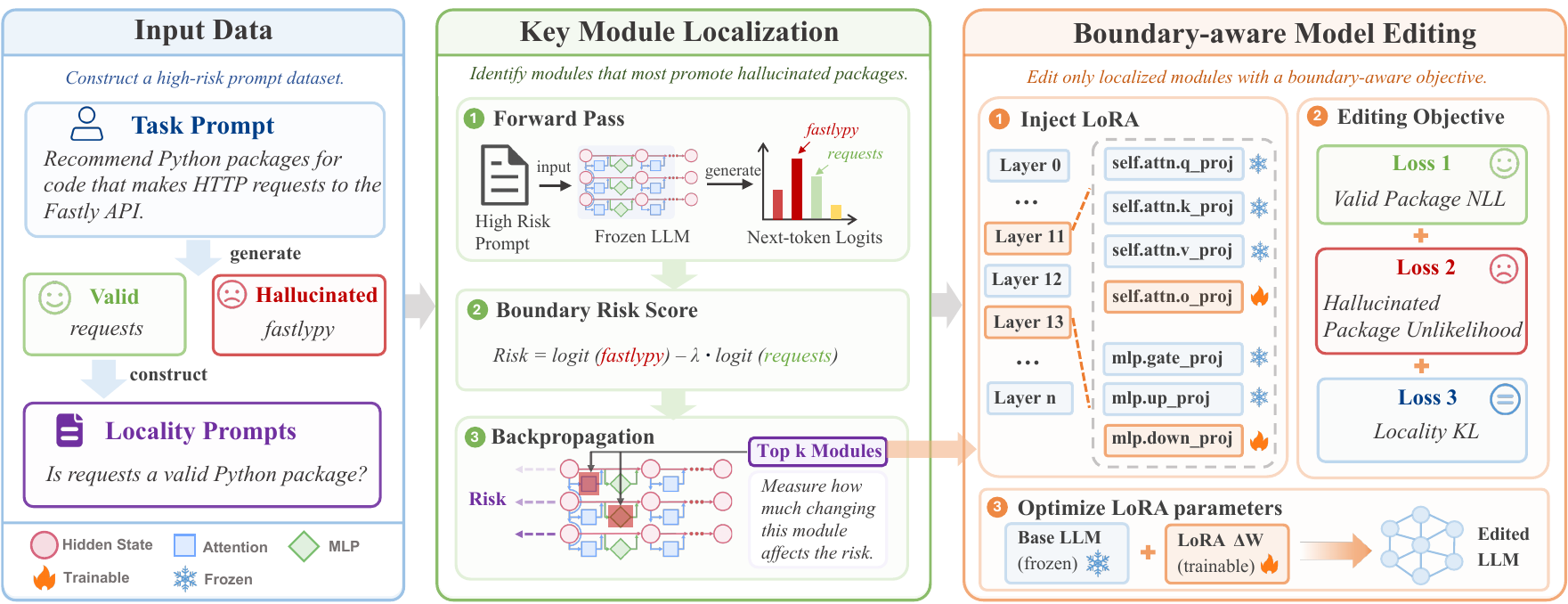}
\caption{Overview of \appname. \ding{182} (left) BOUND constructs edit cases by identifying valid packages, hallucinated packages, and building locality prompts from high-risk task prompts. 
\ding{183} (middle) BOUND computes a boundary risk score and backpropagates it through the frozen LLM to locate modules that most affect the package-validity boundary. 
\ding{184} (right) BOUND injects LoRA into the localized modules and optimizes them with 
complementary objectives to reinforce valid packages, suppress hallucinated packages, and preserve locality behavior.}
\label{fig:bound-approach}
\vspace{-0.3cm}
\end{figure*}

Given a set of high-risk package recommendation prompts, \appname aims to edit the model to reduce hallucinated package names while preserving valid package recommendations. 
We formulate this goal as refining the model's package-validity boundary, i.e., its ability to distinguish valid packages from hallucinated package names under a given task context.
As shown in Figure~\ref{fig:bound-approach}, \appname consists of two editing components. First, key module localization identifies modules that consistently contribute to hallucinated package generation across multiple edit cases. Second, model editing injects lightweight LoRA parameters into the localized modules and optimizes them with a boundary-aware objective that reinforces valid packages, suppresses hallucinated packages, and preserves locality behavior. The following subsections describe these two components in detail.

\subsection{Key Module Localization}
The objective of this stage is to localize where model edits should be applied. Rather than updating all model parameters, \appname identifies modules that are most sensitive to a package-validity boundary risk score.

An autoregressive transformer-based LLM $M_{\theta}$ typically consists of an embedding layer followed by a stack of $L$ decoder layers $\{L_1,L_2,\ldots,L_L\}$.
Each decoder layer contains a self-attention block and an MLP block.
Prior knowledge editing methods primarily focus on MLP modules, based on evidence that feed-forward modules play an important role in storing and updating factual associations~\cite{geva2021transformer,dai2022knowledge,meng2022rome,meng2023memit}.
However, package hallucination is not merely a factual knowledge storage problem.
Package recommendation requires the model to understand the task context, identify useful dependencies, and distinguish valid package names from non-existent names.
Therefore, package-validity boundary errors may involve not only MLP modules that affect vocabulary-level prediction, but also attention modules that route task-relevant contextual information.

To explicitly define the localization space, consider the $\ell$-th decoder layer. Given the hidden state $h_{\ell}$, the layer can be written as:
\begin{equation}
\begin{aligned}
    \tilde{h}_{\ell}
    &=
    h_{\ell}
    +
    \underbrace{
    W_{\ell}^{o}
    \operatorname{Attn}
    \left(
    W_{\ell}^{q} h_{\ell},
    W_{\ell}^{k} h_{\ell},
    W_{\ell}^{v} h_{\ell}
    \right)
    }_{\text{Attention block}}, \\
    h_{\ell+1}
    &=
    \tilde{h}_{\ell}
    +
    \underbrace{
    W_{\ell}^{d}
    \left(
    \sigma(W_{\ell}^{g}\tilde{h}_{\ell})
    \odot
    W_{\ell}^{u}\tilde{h}_{\ell}
    \right)
    }_{\text{MLP block}} .
\end{aligned}
\end{equation}
For clarity, we omit layer normalization and other implementation-specific details.
Here, $W_{\ell}^{q}$, $W_{\ell}^{k}$, $W_{\ell}^{v}$, and $W_{\ell}^{o}$ denote the query, key, value, and output projections in the attention block, respectively. Similarly, $W_{\ell}^{g}$, $W_{\ell}^{u}$, and $W_{\ell}^{d}$ denote the gate, up, and down projections in the MLP block, respectively. \appname treats these projection matrices as candidate modules:
\begin{equation}
    \mathcal{M}_{\ell}
    =
    \left\{
    W_{\ell}^{q},
    W_{\ell}^{k},
    W_{\ell}^{v},
    W_{\ell}^{o},
    W_{\ell}^{g},
    W_{\ell}^{u},
    W_{\ell}^{d}
    \right\}.
\end{equation}

Across all decoder layers, the complete candidate module space is defined as
\begin{equation}
    \mathcal{M}
    =
    \bigcup_{\ell=1}^{L}
    \mathcal{M}_{\ell}.
\end{equation}

During localization, the base model remains frozen. The weights of candidate modules are temporarily marked as gradient-bearing only for sensitivity measurement, and no parameter update is performed in this stage.
Given a set of high-risk package recommendation prompts ($\mathcal{E}_{\mathrm{loc}}$) for localization, \appname follows three steps to identify modules that are most sensitive to package-validity boundary risk:

\noindent\ding{182} \textbf{Forward Pass.}
For each edit case $e_i=(x_i,G_i,H_i)$ in $\mathcal{E}_{\mathrm{loc}}$, \appname feeds the package recommendation prompt $x_i$ into the frozen model $M_{\theta}$ and performs a forward pass to assess the model’s preference among package name candidates at the beginning of the answer. Specifically, \appname extracts the next-token logits:
\begin{equation}
    z_i = M_{\theta}(x_i)_{\mathrm{next}},
\end{equation}
where $z_i \in \mathbb{R}^{|\mathcal{V}|}$ denotes the logit vector over the model vocabulary $\mathcal{V}$.
Since the model predicts over tokens rather than package name strings, \appname maps $G_i$ and $H_i$ to token-level candidate sets $G_i^{\mathrm{tok}}$ and $H_i^{\mathrm{tok}}$ using the tokenizer, and extracts their logits from $z_i$ as $z_i[G_i^{\mathrm{tok}}]$ and $z_i[H_i^{\mathrm{tok}}]$.

\noindent\ding{183} \textbf{Boundary Risk Score.}
\appname then constructs a boundary risk score by contrasting hallucinated package candidates with valid package candidates:
\begin{equation}
R_i =
\operatorname{LSE}\left(z_i[H_i^{\mathrm{tok}}]\right)
-
\lambda_a
\operatorname{LSE}\left(z_i[G_i^{\mathrm{tok}}]\right),
\end{equation}
where $\operatorname{LSE}(\cdot)$ computes a smooth aggregate score over a set of token logits, and $\lambda_a$ controls the weight of the valid package anchor term. The first term quantifies the model’s preference for hallucinated package names, whereas the second term serves as a valid package anchor that offsets this preference. Consequently, a larger value of $R_i$ indicates a stronger model preference for hallucinated packages relative to valid packages under the current task context.

\noindent\ding{184} \textbf{Backpropagation.}
After computing the boundary risk score, \appname backpropagates $R_i$ through the frozen model to estimate the sensitivity of each candidate module to the package-validity boundary. For a candidate module $m \in \mathcal{M}$ with weight matrix $W_m$, \appname computes a saliency score:
\begin{equation}
s_i(m)
=
\operatorname{Mean}
\left(
\left|
\nabla_{W_m} R_i
\odot
W_m
\right|
\right),
\end{equation}
where $\nabla_{W_m} R_i$ denotes the gradient of $R_i$ with respect to $W_m$, and $\odot$ denotes element-wise multiplication.
This score combines the gradient signal with the current weight values and summarizes their absolute effect at the module level.
A higher score indicates that module $m$ has a stronger influence on the package-validity boundary and is therefore a more promising editing target.
\appname then averages the saliency scores over all localization cases:
\begin{equation}
S(m)
=
\frac{1}{|\mathcal{E}_{\mathrm{loc}}|}
\sum_{e_i \in \mathcal{E}_{\mathrm{loc}}}
s_i(m),
\end{equation}
\appname ranks candidate modules by $S(m)$ and selects the top modules as the localized module set $\mathcal{S}$ for subsequent editing.

\subsection{Boundary-aware Model Editing}
After localizing the key modules, \appname updates only the localized module set $\mathcal{S}$ instead of the entire model. This stage refines the model's package-validity boundary by suppressing hallucinated package names while preserving valid package recommendations. \appname performs the editing process in three steps:

\noindent\ding{182} \textbf{Injecting LoRA.}
Given the localized module set $\mathcal{S}$, \appname injects lightweight LoRA adapters into each selected linear module~\cite{hu2022lora}. For a module $m \in \mathcal{S}$, let $h_m$ denote the input hidden representation and $o_m$ denote the module output. The original module computes
\begin{equation}
o_m = W_m h_m,
\end{equation}
where $W_m$ is the weight matrix. After LoRA injection, the module output is modified as follows:
\begin{equation}
o_m^{\prime}
=
W_m h_m
+
\frac{\alpha}{r}
B_m A_m \operatorname{Dropout}(h_m),
\end{equation}
where $A_m$ and $B_m$ are trainable low-rank matrices, $r$ denotes the LoRA rank, and $\alpha$ is a scaling factor. During editing, the original weight matrix $W_m$ remains frozen, and only the injected LoRA parameters are optimized. The LoRA branch is initialized to produce zero output, so the edited model initially preserves the behavior of the original model. We denote the collection of all LoRA updates on localized modules as $\Delta$.

\noindent\ding{183} \textbf{Editing Objective.}
For each edit case $e_i=(x_i,G_i,H_i)$, \appname jointly optimizes three complementary objectives:

\begin{itemize}[leftmargin=0.5cm]

\item \textbf{\textit{Valid Package NLL.}} \appname uses a negative log-likelihood (NLL) loss on the formatted valid package answer $y_i$ as positive supervision.
We construct the target $y_i$ by formatting the valid packages $G_i$, but it is not assumed to include all correct packages for the task.
Instead, it encourages the edited model to keep valid package names under the prompt $x_i$:
\begin{equation}
    \mathcal{L}_{\mathrm{valid}}^{(i)}
    =
    -
    \frac{1}{|y_i|}
    \sum_{t=1}^{|y_i|}
    \log P_{\theta+\Delta}
    \left(
    y_{i,t}
    \mid
    x_i,y_{i,<t}
    \right),
\end{equation}
where $y_{i,t}$ is the $t$-th token of $y_i$, $y_{i,<t}$ is its preceding target context, and $P_{\theta+\Delta}$ denotes the edited model. Minimizing this loss preserves valid package generation and prevents the model from reducing hallucinations by simply suppressing package outputs.

\item \textbf{\textit{Hallucinated Package Unlikelihood.}} \appname uses an unlikelihood loss~\cite{welleck2019neural} to explicitly suppress hallucinated package names in $H_i$. For each hallucinated package $h \in H_i$, \appname treats it as a candidate continuation after the answer prefix $\pi$, where $\pi$ is the prefix used to start the package list. 
Let $p_{\theta+\Delta}(h \mid x_i,\pi)$ denote the normalized sequence probability of generating $h$ under the edited model. 
\appname penalizes hallucinated packages that remain after editing:
\begin{equation}
    \mathcal{L}_{\mathrm{hall}}^{(i)}
    =
    -
    \frac{1}{|H_i|}
    \sum_{h \in H_i}
    \log
    \left(
    1 -
    p_{\theta+\Delta}(h \mid x_i,\pi)
    \right),
\end{equation}
minimizing this objective pushes the model to reduce the probability of hallucinated package names.

\item \textbf{\textit{Locality KL.}} \appname uses a locality loss based on Kullback-Leibler (KL) divergence to reduce unintended changes caused by editing. For each edit case, we construct a locality set $C_i$ from two sources. First, \appname derives package-related locality prompts from the observed valid packages in $G_i$ to preserve the model’s behavior around valid package knowledge. Second, \appname includes the original task prompt $x_i$ to prevent the edited model from overfitting to the formatted target answer $y_i$. For each locality prompt $c \in C_i$, \appname encourages the edited model to preserve the next token distribution of the base model:
\begin{equation}
    \mathcal{L}_{\mathrm{loc}}^{(i)}
    =
    \frac{1}{|C_i|}
    \sum_{c \in C_i}
    D_{\mathrm{KL}}
    \left(
    P_{\theta}(\cdot \mid c)
    \,\|\, 
    P_{\theta+\Delta}(\cdot \mid c)
    \right),
\end{equation}
Here, $P_{\theta}(\cdot \mid c)$ and $P_{\theta+\Delta}(\cdot \mid c)$ denote the next token distributions of the original and edited models under prompt $c$. 
Minimizing this loss constrains the edited model and reduces distribution drift on valid package contexts and the original task context.
\end{itemize}

Overall, the editing objective is
\begin{equation}
    \mathcal{L}_{\mathrm{BOUND}}
    =
    \frac{1}{|\mathcal{E}_{\mathrm{edit}}|}
    \sum_{e_i \in \mathcal{E}_{\mathrm{edit}}}
    \left(
    \lambda_v \mathcal{L}_{\mathrm{valid}}^{(i)}
    +
    \lambda_h \mathcal{L}_{\mathrm{hall}}^{(i)}
    +
    \lambda_l \mathcal{L}_{\mathrm{loc}}^{(i)}
    \right),
\end{equation}
where $\lambda_v$, $\lambda_h$, and $\lambda_l$ control the strengths of valid package reinforcement, hallucinated package suppression, and locality preservation, respectively.

\noindent\ding{184} \textbf{Optimizing LoRA parameters.}
\appname minimizes $\mathcal{L}_{\mathrm{BOUND}}$ by updating the LoRA parameters injected into the localized module set $\mathcal{S}$, while keeping the base parameters $\theta$ fixed. 
After optimization, the learned LoRA updates form the edit $\Delta$, and the edited model is denoted as $M_{\theta+\Delta}$. 
By combining localized updates with valid package reinforcement, hallucinated package suppression, and locality preservation, \appname refines the model's package-validity boundary instead of merely fitting valid package targets.

\section{Experimental Setup}
\label{sec:experimentalsetup}
\subsection{Dataset}
\label{sec:dataset}
We construct a high-risk prompt set for each evaluated model and further split it into edit and unseen evaluation subsets. The dataset construction process contains three steps:

\noindent\ding{182} \textbf{Prompt source.}
We follow the LLM-generated prompt dataset released by Spracklen et al.~\cite{spracklen2025we} and use its recent Python subset. 
The prompts are derived from descriptions of popular Python packages and cover diverse package-related programming tasks.

\noindent\ding{183} \textbf{High-risk prompt construction.}
Since package hallucination behavior varies across models, we construct a separate high-risk prompt set for each evaluated model.
For each prompt $x$, we query the model five times in the package recommendation setting, extract the recommended packages, and check their validity using PyPI metadata~\cite{pypi}. 
A package is valid only if it exists on PyPI and has at least one release before the model's knowledge cutoff. Otherwise, it is classified as a hallucinated package.
We then compute the sample-level hallucination rate (Sample-HR) of each prompt over the five generations, as defined in Equation~\ref{eqa:sample-HR}.
We retain a prompt as a high-risk prompt if Sample-HR $>0.5$, focusing our study on prompts that consistently trigger package hallucinations.

\noindent\ding{184} \textbf{Dataset split.}
Table~\ref{tab:dataset_statistics} summarizes the resulting high-risk prompt sets. 
For each model, we split high-risk prompts into four edit subsets and one unseen evaluation subset. 
Each edit subset contains 50 prompts and is used independently for editing, reducing the effect of randomness in edit case selection.
For each edit prompt $x_i$, we construct an edit case $e_i=(x_i,G_i,H_i)$, where $G_i$ and $H_i$ are the extracted valid and hallucinated packages from the five generations.
The remaining high-risk prompts are used to evaluate whether the package-validity boundary learned through editing can generalize to unseen prompts.

\begin{table}[t]
\centering
\caption{Statistics of high-risk prompt sets for the evaluated models.}
\label{tab:dataset_statistics}
\begin{tabular}{lccc}
\toprule
Model & High-risk Prompt & Edit Prompt & Evaluation Prompt \\
\midrule
DeepSeekCoder & 741   & 4 $\times$ 50 & 541 \\
Qwen3         & 609   & 4 $\times$ 50 & 409 \\
Llama-3.1     & 1,134 & 4 $\times$ 50 & 934 \\
\bottomrule
\end{tabular}
\vspace{-0.3cm}
\end{table}

\subsection{Baselines}
To evaluate the effectiveness of \appname, we compare it with five baselines, including two package hallucination mitigation methods and three model editing methods. 
For a fair comparison, all trainable baselines use the same high-risk edit cases as \appname.

\noindent\textbf{Full-FT~\cite{spracklen2025we}} is a direct model adaptation baseline for mitigating package hallucinations.
We construct supervised examples by pairing each package recommendation prompt $x_i$ with its valid package answer $y_i$, and fine-tune the model to increase the likelihood of valid package recommendations.

\noindent\textbf{Self-Refinement~\cite{spracklen2025we}} is an inference-time baseline that iteratively verifies and refines package recommendations. 
In each round, the model generates packages, checks their validity, adds invalid packages to an avoidance list, and regenerates the answer. 
We repeat this process for up to five rounds.

\noindent\textbf{ROME~\cite{meng2022rome}} is a localized knowledge editing method that applies rank-one parameter updates to edit factual associations. 
We adapt ROME by using the package recommendation prompt $x_i$ as the editing context and the valid package answer $y_i$ as the target output.

\noindent\textbf{MEMIT~\cite{meng2023memit}} extends ROME to edit multiple factual associations in a batch. 
We use the same edit cases to evaluate whether batch factual editing can mitigate package hallucination at a larger edit scale.

\noindent\textbf{DINM~\cite{wang2024detoxifying}} is a behavior editing method originally designed for tasks such as detoxification. 
We adapt DINM by treating $y_i$ as the desired behavior, and evaluate whether behavior editing can suppress hallucinated package recommendations while preserving valid packages.

\subsection{Evaluation Metrics}
We evaluate each method from two perspectives: package hallucination and valid package preservation. For each generated answer $a$, we extract the set of recommended packages $P(a)$ and classify each package using the cutoff-aware PyPI validation procedure described in Section~\ref{sec:dataset}. Let $G(a)$ and $H(a)$ denote the sets of valid and hallucinated packages extracted from $a$, respectively.

\noindent\textbf{Sample-HR:} Sample-level hallucination rate measures the fraction of generated answers that contain at least one hallucinated package:
\begin{equation}
\label{eqa:sample-HR}
\mathrm{Sample\mbox{-}HR}
=
\frac{1}{N}
\sum_{i=1}^{N}
\mathbb{I}\left(|H(a_i)| > 0\right),
\end{equation}
where $N$ is the number of generated answers. Sample-HR reflects how frequently users are exposed to hallucinated package recommendations.

\noindent\textbf{Package-HR:} Package-level hallucination rate measures the fraction of hallucinated packages among extracted packages:
\begin{equation}
\mathrm{Package\mbox{-}HR}
=
\frac{
\sum_{i=1}^{N} |H(a_i)|
}{
\sum_{i=1}^{N} \left(|G(a_i)| + |H(a_i)|\right)
}.
\end{equation}
This metric reflects the overall hallucination level among generated package names. If no package is extracted from any generated answer, we define Package-HR as 0.

\noindent\textbf{Valid-Rate:} Valid-Rate measures the fraction of generated answers that contain at least one valid package:
\begin{equation}
\mathrm{Valid\mbox{-}Rate}
=
\frac{1}{N}
\sum_{i=1}^{N}
\mathbb{I}\left(|G(a_i)| > 0\right).
\end{equation}
Here, a valid package means a real package that exists in the package ecosystem. It does not imply that the package is semantically correct for the given task. 
Valid-Rate complements the hallucination metrics, since a method may reduce hallucinations by suppressing package generation.

\subsection{Experimental Setting}

\noindent\textbf{Models.}
We evaluate on three open-source LLMs: \texttt{deepseek-coder-6.7b-instruct}~\cite{guo2024deepseek}, \texttt{Qwen3-8B}
\cite{yang2025qwen3}, and \texttt{Llama-3.1-8B-Instruct}~\cite{grattafiori2024llama}. All experiments use PyTorch and HuggingFace Transformers, with model weights loaded in \texttt{bfloat16} precision.

\noindent\textbf{Editing Hyperparameters.}
\appname selects the top 5 modules during localization. 
During editing, it uses a LoRA rank of $r=8$, scaling factor $\alpha=16$, dropout rate of 0.05, batch size of 1, learning rate of $2\times10^{-4}$, and three training epochs. The loss weights $\lambda_v$, $\lambda_h$, and $\lambda_l$ are set to 1.0, 0.15, and 0.03, respectively. For a fair comparison, all trainable baselines use the same edit cases as \appname, while method-specific hyperparameters follow their default settings.

\noindent\textbf{Generation Settings.}
For all evaluations, each prompt is sampled five times using \texttt{temperature} = 0.7 and \texttt{top\_p} = 0.9. The maximum generation length is set to 128 tokens for package recommendation and \texttt{pip install} recommendation tasks, and 2,048 tokens for code generation.

\noindent\textbf{Hardware.}
All experiments are conducted on a Linux server running Ubuntu 22.04 LTS, equipped with two Intel Xeon Platinum 8358P CPUs (64 cores total @ 2.60 GHz), 1 TB RAM, and eight NVIDIA A100-SXM4 GPUs (80~GB memory).

\section{Results and Analysis}
\label{sec:results}
\begin{table*}[t]
\centering
\caption{Effectiveness comparison in the package recommendation setting. We report results on both Edit50 prompts and unseen evaluation prompts. Lower Sample-HR and Package-HR are better, while higher Valid-Rate is better.}
\label{tab:main_results_three_models}
\resizebox{\textwidth}{!}{
\begin{tabular}{llccc|ccc}
\toprule
\multirow{2}{*}{\textbf{Model}} & \multirow{2}{*}{\textbf{Method}}
& \multicolumn{3}{c|}{\textbf{Edit50 Prompts}}
& \multicolumn{3}{c}{\textbf{Unseen Evaluation Prompts}} \\
\cmidrule(lr){3-5} \cmidrule(lr){6-8}
& & Sample-HR $\downarrow$ & Package-HR $\downarrow$ & Valid-Rate $\uparrow$
  & Sample-HR $\downarrow$ & Package-HR $\downarrow$ & Valid-Rate $\uparrow$ \\
\midrule

\multirow{7}{*}{DeepSeekCoder}
& Base            & 0.8140 & 0.4091 & 0.8270 & 0.8333 & 0.4414 & 0.7978 \\
& \cellcolor{boundRow}\textbf{\appname} & \cellcolor{boundRow}0.1980 & \cellcolor{boundRow}0.0745 & \cellcolor{boundRow}0.9480 & \cellcolor{boundRow}0.3712 & \cellcolor{boundRow}0.1706 & \cellcolor{boundRow}0.8909 \\
& ROME            & 0.1820 & 0.0916 & 0.2120 & 0.1879 & 0.1085 & 0.1978 \\
& MEMIT           & 0.7560 & 0.4049 & 0.8160 & 0.7503 & 0.4114 & 0.8001 \\
& DINM            & 0.1540 & 0.1327 & 0.4000 & 0.2805 & 0.2366 & 0.3707 \\
& Full-FT         & 0.0580 & 0.0188 & 0.9500 & 0.4814 & 0.2141 & 0.8991 \\
& Self-Refinement & 0.5180 & 0.2263 & 0.8360 & --     & --     & --     \\
\midrule

\multirow{7}{*}{Llama-3.1}
& Base            & 0.8320 & 0.4308 & 0.8150 & 0.8368 & 0.4295 & 0.8182 \\
& \cellcolor{boundRow}\textbf{\appname} & \cellcolor{boundRow}0.3180 & \cellcolor{boundRow}0.1419 & \cellcolor{boundRow}0.8690 & \cellcolor{boundRow}0.4033 & \cellcolor{boundRow}0.1955 & \cellcolor{boundRow}0.8353 \\
& ROME            & 0.1590 & 0.4297 & 0.0110 & 0.1587 & 0.4325 & 0.0147 \\
& MEMIT           & 0.7340 & 0.3899 & 0.8310 & 0.7389 & 0.3870 & 0.8304 \\
& DINM            & 0.0320 & 0.0423 & 0.2710 & 0.0387 & 0.0590 & 0.2591 \\
& Full-FT         & 0.5110 & 0.3064 & 0.8660 & 0.5592 & 0.3546 & 0.8265 \\
& Self-Refinement & 0.7120 & 0.3897 & 0.7740 & --     & --     & --     \\
\midrule

\multirow{7}{*}{Qwen3}
& Base            & 0.9040 & 0.4991 & 0.7620 & 0.9017 & 0.4555 & 0.7990 \\
& \cellcolor{boundRow}\textbf{\appname} & \cellcolor{boundRow}0.0990 & \cellcolor{boundRow}0.0526 & \cellcolor{boundRow}0.7410 & \cellcolor{boundRow}0.1595 & \cellcolor{boundRow}0.0927 & \cellcolor{boundRow}0.7434 \\
& ROME            & 0.3760 & 0.1879 & 0.6640 & 0.6069 & 0.4036 & 0.4475 \\
& MEMIT           & 0.8440 & 0.4757 & 0.7520 & 0.8318 & 0.4423 & 0.7883 \\
& DINM            & 0.7670 & 0.4372 & 0.7480 & 0.7545 & 0.4124 & 0.7791 \\
& Full-FT         & 0.1460 & 0.0895 & 0.7780 & 0.3341 & 0.2572 & 0.6417 \\
& Self-Refinement & 0.5650 & 0.2668 & 0.6860 & --     & --     & --     \\

\bottomrule
\end{tabular}
}
\vspace{-0.3cm}
\end{table*}

In this section, we aim to answer the following three research questions:

\noindent\textbf{RQ1 (Effectiveness).} How effective is \appname compared to package hallucination mitigation and model editing methods?

\noindent\textbf{RQ2 (Generalization).} Can \appname generalize from package recommendation to other package-related tasks?

\noindent\textbf{RQ3 (Ablation Study).} How does each component of \appname contribute to its overall effectiveness?

\subsection{RQ1: Effectiveness}
To answer RQ1, we evaluate \appname and all baselines in the package recommendation setting. We also report the unedited model as \textbf{Base}. 
Following the dataset split in Table~\ref{tab:dataset_statistics}, all trainable methods are trained using the same four edit folds, each containing 50 prompts. For each method, we report the average performance across the four folds on both the edit set and the unseen evaluation set. Table~\ref{tab:main_results_three_models} presents the results. 

\noindent\faThumbsUp~\textbf{Effectiveness of \appname:}
\appname substantially reduces package hallucinations across all three models while preserving valid package recommendations. On the edit prompts, \appname reduces the average Sample-HR from 0.8500 to 0.2050 (-75.9\%) and the average Package-HR from 0.4463 to 0.0897 (-79.9\%). Meanwhile, the average Valid-Rate increases from 0.8013 to 0.8527 (+6.4\%), indicating that \appname suppresses hallucinated packages without sacrificing valid package generation.
The improvements further generalize to unseen prompts that are not used during editing.
On the unseen evaluation prompts, \appname reduces the average Sample-HR from 0.8573 to 0.3113 (-63.7\%) and the average Package-HR from 0.4421 to 0.1529 (-65.4\%). These consistent gains suggest that \appname learns a transferable package-validity boundary rather than simply memorizing the edited prompts.

\noindent\faThumbsUp~\textbf{Comparison with Baselines:}
Compared with model editing baselines, \appname achieves a better balance between hallucination reduction and valid package preservation.
Although ROME and DINM sometimes achieve very low Sample-HR or Package-HR, these gains are often accompanied by substantial drops in Valid-Rate. For example, on Llama-3.1, ROME reduces the Edit50 Sample-HR to 0.1590, but its Valid-Rate drops to only 0.0110. Similar trends can be observed for DINM across multiple models.
This suggests that these methods may reduce hallucinations by suppressing package recommendation behavior itself. 
In contrast, \appname reduces hallucinations while maintaining a high Valid-Rate, indicating that it better separates valid packages from hallucinated ones.
\appname also generalizes better than package hallucination mitigation baselines.
Although Full-FT performs competitively on some edit sets, its improvements transfer less consistently to unseen prompts.
On the unseen evaluation prompts, \appname reduces Package-HR by 65.4\% on average across the three models, compared with 37.8\% for Full-FT.
Self-Refinement achieves moderate hallucination reductions, but its gains are smaller than those of \appname (29.6\% vs. 75.9\% Sample-HR reduction on Edit50).
Furthermore, Self-Refinement can be applied to any prompt at inference time. 
However, since our unseen evaluation focuses on the generalization of the edited model, Self-Refinement is not included in this comparison.

\begin{center}
    \resizebox{\linewidth}{!}{
\begin{tabular}{l!{\vrule width 1pt}p{0.9\columnwidth}}
    \makecell{{\LARGE \faLightbulbO}}  &\textbf{Answer to RQ1:}
        \appname achieves the best overall balance between hallucinated package mitigation and valid package preservation. On edit prompts, it reduces Sample-HR and Package-HR by 75.9\% and 79.9\%, while improving Valid-Rate by 6.4\%. These improvements also generalize to unseen prompts, reducing Sample-HR and Package-HR by 63.7\% and 65.4\%.\\
\end{tabular}}
\end{center}

\subsection{RQ2: Generalization}
\begin{table*}[t]
\centering
\caption{Cross-task generalization results on code generation and \texttt{pip install} recommendation.}
\label{tab:rq2_cross_task}
\footnotesize
\resizebox{.95\textwidth}{!}{
\begin{tabular}{llccc|ccc}
\toprule
\multirow{2}{*}{\textbf{Model}} & \multirow{2}{*}{\textbf{Method}}
& \multicolumn{3}{c|}{\textbf{Code Generation}}
& \multicolumn{3}{c}{\textbf{\texttt{pip install} Recommendation}} \\
\cmidrule(lr){3-5} \cmidrule(lr){6-8}
& & Sample-HR $\downarrow$ & Package-HR $\downarrow$ & Valid-Rate $\uparrow$
  & Sample-HR $\downarrow$ & Package-HR $\downarrow$ & Valid-Rate $\uparrow$ \\
\midrule

\multirow{6}{*}{DeepSeekCoder}
& Base    & 0.3420 & 0.2643 & 0.6420 & 0.4340 & 0.2986 & 0.6320 \\
& \cellcolor{boundRow}\textbf{\appname} 
          & \cellcolor{boundRow}0.3260 & \cellcolor{boundRow}0.2542 & \cellcolor{boundRow}0.6415
          & \cellcolor{boundRow}0.3375 & \cellcolor{boundRow}0.1987 & \cellcolor{boundRow}0.7690 \\
& DINM    & 0.3105 & 0.2714 & 0.5655 & 0.2070 & 0.2352 & 0.3985 \\
& MEMIT   & 0.3550 & 0.2758 & 0.6215 & 0.4150 & 0.3017 & 0.6130 \\
& ROME    & 0.0850 & 0.0671 & 0.1510 & 0.1010 & 0.0669 & 0.1515 \\
& Full-FT & 0.3540 & 0.2729 & 0.6385 & 0.3845 & 0.2547 & 0.6275 \\
\midrule

\multirow{6}{*}{Qwen3}
& Base    & 0.3700 & 0.2785 & 0.5500 & 0.4820 & 0.3569 & 0.6040 \\
& \cellcolor{boundRow}\textbf{\appname} 
          & \cellcolor{boundRow}0.3385 & \cellcolor{boundRow}0.2237 & \cellcolor{boundRow}0.5815
          & \cellcolor{boundRow}0.2400 & \cellcolor{boundRow}0.2122 & \cellcolor{boundRow}0.6030 \\
& DINM    & 0.3230 & 0.2682 & 0.4820 & 0.5015 & 0.3704 & 0.5945 \\
& MEMIT   & 0.3640 & 0.2829 & 0.5200 & 0.4910 & 0.3608 & 0.6050 \\
& ROME    & 0.0935 & 0.0715 & 0.1325 & 0.1215 & 0.0871 & 0.1545 \\
& Full-FT & 0.0930 & 0.1094 & 0.3200 & 0.0000 & 0.0000 & 0.0005 \\
\midrule

\multirow{6}{*}{Llama-3.1}
& Base    & 0.3380 & 0.2425 & 0.6940 & 0.5760 & 0.3452 & 0.6600 \\
& \cellcolor{boundRow}\textbf{\appname} 
          & \cellcolor{boundRow}0.3085 & \cellcolor{boundRow}0.2070 & \cellcolor{boundRow}0.7165
          & \cellcolor{boundRow}0.3825 & \cellcolor{boundRow}0.2492 & \cellcolor{boundRow}0.7675 \\
& DINM    & 0.0095 & 0.1352 & 0.0115 & 0.0005 & 0.2500 & 0.0000 \\
& MEMIT   & 0.3355 & 0.2319 & 0.6890 & 0.5650 & 0.3348 & 0.6785 \\
& ROME    & 0.0000 & 0.0000 & 0.0000 & 0.0000 & 0.0000 & 0.0000 \\
& Full-FT & 0.0000 & 0.0000 & 0.0005 & 0.0000 & 0.0000 & 0.0000 \\

\bottomrule
\end{tabular}
}
\vspace{-0.3cm}
\end{table*}
RQ1 shows the effectiveness of \appname in the package recommendation task. We further investigate whether the package-validity boundary knowledge learned from package recommendation can transfer to other package-related tasks.
We evaluate two representative tasks: code generation and \texttt{pip install} recommendation.
For code generation, the edited model first generates Python code for a given task.
We then extract the \texttt{import} modules and use the corresponding unedited base model to map them to PyPI package names.
This design separates package name mapping from code generation quality, allowing the evaluation to focus on how editing affects dependencies in generated code.
For \texttt{pip install} recommendation, we give the model the task prompt and ask it to generate \texttt{pip install} commands directly. 
We then extract package names from the generated commands for validity checking.
To control evaluation cost, we randomly sample 100 prompts from the unseen evaluation prompt set. 
Following the RQ1 setup, we evaluate the four edited models obtained from the four edit folds and report their average results.
We exclude Self-Refinement because it is an inference-time mitigation method and does not produce edited model parameters that can transfer across tasks.
Table~\ref{tab:rq2_cross_task} presents the results.

\noindent\faHandORight~\textbf{Cross-Task Generalization:}
The edited package-validity boundary transfers from package recommendation to both code generation and \texttt{pip install} recommendation.
In code generation, \appname reduces the average Sample-HR and Package-HR by 7.3\% and 12.8\%, respectively, while improving the average Valid-Rate by 2.8\%.
This indicates that editing package recommendation behavior can also reduce hallucinated dependencies in generated code.
However, the gains are smaller than those in the package recommendation setting.
One possible reason is that package hallucinations in code generation are often introduced through hallucinated module imports, making the effect of package-validity boundary editing less direct than in explicit package recommendation.
\appname shows stronger transfer in the \texttt{pip install} recommendation setting.
Across the three models, it reduces Sample-HR and Package-HR by 35.7\% and 34.0\%, while improving Valid-Rate by 12.8\%.
This suggests that the edited package-validity boundary generalizes well to explicit package installation tasks, where the model still needs to determine whether a package name is valid.

\noindent\faHandORight~\textbf{Comparison with Baselines:}
ROME, DINM, and Full-FT sometimes achieve very low Sample-HR or Package-HR, but often at the cost of valid package generation. For example, on Llama-3.1 in the \texttt{pip install} setting, ROME and Full-FT both achieve zero Sample-HR and Package-HR, while their Valid-Rate also drops to zero. 
This suggests that these methods may damage the model's basic ability to perform package-related tasks, leading to fewer package outputs rather than reducing hallucinations.
In contrast, \appname consistently reduces hallucinations while maintaining a high Valid-Rate, indicating that it better preserves valid package behavior.
MEMIT preserves Valid-Rate more effectively than ROME and DINM, but its hallucination reduction is limited. In both code generation and \texttt{pip install} recommendation, its Sample-HR and Package-HR remain close to those of the Base model. This suggests that MEMIT preserves the original model behavior, but has limited ability to transfer the edited package-validity boundary knowledge across tasks.
Overall, these results indicate that directly applying existing methods is insufficient for cross-task package hallucination mitigation, whereas \appname provides a more reliable package-validity boundary editing approach.

\begin{center}
    \resizebox{\linewidth}{!}{
\begin{tabular}{l!{\vrule width 1pt}p{0.9\columnwidth}}
    \makecell{{\LARGE \faLightbulbO}}  &\textbf{Answer to RQ2:}
        \appname transfers the package-validity boundary learned from package recommendation to other package-related tasks. It reduces Package-HR by 12.8\% in code generation and by 34.0\% in \texttt{pip install} recommendation. These results suggest that \appname learns a transferable distinction between valid and hallucinated packages, rather than merely fitting the edited recommendation prompts.\\
\end{tabular}}
\end{center}

\subsection{RQ3: Ablation Study}
\begin{table*}[t]
\centering
\caption{Ablation study of \appname across three models.}
\label{tab:rq3_ablation}
\setlength{\tabcolsep}{4pt}
\renewcommand{\arraystretch}{0.92}
\resizebox{\textwidth}{!}{
\begin{tabular}{lccc|ccc|ccc}
\toprule
\multirow{2}{*}{\textbf{Setting}}
& \multicolumn{3}{c|}{\textbf{DeepSeekCoder}}
& \multicolumn{3}{c|}{\textbf{Qwen3}}
& \multicolumn{3}{c}{\textbf{Llama-3.1}} \\
\cmidrule(lr){2-4} \cmidrule(lr){5-7} \cmidrule(lr){8-10}
& Sample-HR $\downarrow$ & Package-HR $\downarrow$ & Valid-Rate $\uparrow$
& Sample-HR $\downarrow$ & Package-HR $\downarrow$ & Valid-Rate $\uparrow$
& Sample-HR $\downarrow$ & Package-HR $\downarrow$ & Valid-Rate $\uparrow$ \\
\midrule
\cellcolor{boundRow}\textbf{\appname}
& \cellcolor{boundRow}0.3620 & \cellcolor{boundRow}0.1618 & \cellcolor{boundRow}0.8835
& \cellcolor{boundRow}0.1300 & \cellcolor{boundRow}0.0702 & \cellcolor{boundRow}0.7265
& \cellcolor{boundRow}0.4700 & \cellcolor{boundRow}0.2204 & \cellcolor{boundRow}0.8645 \\

w/o Localization
& 0.4110 & 0.2356 & 0.7590
& 0.2055 & 0.1548 & 0.5890
& 0.5215 & 0.2555 & 0.8420 \\

w/o Valid-NLL
& 0.0200 & 0.0102 & 0.2140
& 0.0000 & 0.0000 & 0.0000
& 0.0855 & 0.0893 & 0.2395 \\

w/o Hall-UL
& 0.6435 & 0.4728 & 0.6435
& 0.5545 & 0.5706 & 0.3205
& 0.7575 & 0.6505 & 0.3980 \\

w/o Locality-KL
& 0.6420 & 0.5668 & 0.4625
& 0.5065 & 0.5214 & 0.3175
& 0.7120 & 0.6027 & 0.4190 \\
\bottomrule
\end{tabular}
}
\vspace{-0.3cm}
\end{table*}
To investigate the contribution of each component to the effectiveness of \appname, we create the following variants:
\begin{itemize}[leftmargin=*]
    \item \textbf{BOUND\_Full}: The complete version of \appname, as described in Section~\ref{sec:approach}.
    \item \textbf{w/o Localization}: This variant removes module localization and randomly selects the same number of editable modules as \appname. It evaluates whether locating modules related to the package-validity boundary is necessary.
    \item \textbf{w/o Valid-NLL}: This variant removes the valid package NLL loss, which uses valid package answers as positive supervision. It evaluates whether valid package learning is needed to preserve useful recommendations.
    \item \textbf{w/o Hall-UL}: This variant removes the hallucinated package unlikelihood loss and no longer directly suppresses hallucinated packages. It evaluates whether explicit hallucination suppression is necessary.
    \item \textbf{w/o Locality-KL}: This variant removes the locality KL loss, so the edited model is no longer constrained to preserve the original behavior on locality prompts. It evaluates whether locality preservation is needed for stable editing.
\end{itemize}

Following the setup of RQ2, we randomly sample 100 prompts from the unseen prompt set to control evaluation cost. For each variant, we edit the model separately on the four edit sets and report the average results across the four edited models. Table~\ref{tab:rq3_ablation} presents the results.

Overall, the complete version of \appname achieves the best overall effectiveness among all variants, balancing hallucination reduction and valid package preservation.
Across the three models, it obtains an average Sample-HR of 0.3207, Package-HR of 0.1508, and Valid-Rate of 0.8248.
Removing localization increases Sample-HR and Package-HR by 18.3\% and 42.8\%, while reducing Valid-Rate by 11.5\%. 
This result suggests that editing localized modules is more effective than editing randomly selected modules.
The three editing objectives also contribute to this balance. 
Removing Valid-NLL reduces Valid-Rate from 0.8248 to 0.1512 (-81.7\%), indicating that the model tends to suppress package generation instead of preserving valid packages. 
Removing Hall-UL increases Package-HR from 0.1508 to 0.5646 (+274.4\%), showing that suppressing hallucinated package names is necessary for hallucination reduction. 
Removing Locality-KL increases Package-HR to 0.5636 (+273.8\%) and reduces Valid-Rate to 0.3997 (-51.5\%), suggesting that locality preservation helps stabilize editing and prevents harmful behavioral drift.

\begin{center}
    \resizebox{\linewidth}{!}{
\begin{tabular}{l!{\vrule width 1pt}p{0.9\columnwidth}}
    \makecell{{\LARGE \faLightbulbO}}  &\textbf{Answer to RQ3:}
        Each component contributes to the effectiveness of \appname. Module localization identifies modules sensitive to the package-validity boundary, while the three loss terms work together to balance hallucination suppression and valid package preservation.\\
\end{tabular}}
\end{center}

\section{Discussion}
\label{sec:discussion}
\subsection{Cost of \appname}
\begin{table}[t]
\centering
\caption{Average cost of \appname and baselines across models.}
\label{tab:cost_bound}
\footnotesize
\setlength{\tabcolsep}{5pt}
\renewcommand{\arraystretch}{0.95}
\resizebox{\columnwidth}{!}{
\begin{tabular}{lcccc}
\toprule
Method 
& Loc. time (s) 
& Edit time (s) 
& Test time (s) 
& Artifact size \\
\midrule
\cellcolor{boundRow}\textbf{\appname}
& \cellcolor{boundRow}23.81 
& \cellcolor{boundRow}83.09 
& \cellcolor{boundRow}11.30 
& \cellcolor{boundRow}731.03 KB \\

ROME
& 10.19 
& 384.45 
& 10.86 
& 98.01 MB \\

MEMIT
& -- 
& 193.06 
& 13.17 
& 588.01 MB \\

DINM
& 0.92 
& 11.97 
& 12.63 
& 126.68 MB \\

Full-FT
& -- 
& 124.16 
& 15.31 
& 14.27 GB \\

\scriptsize{Self-Refinement}
& -- 
& --  
& 25.12 
& --  \\
\bottomrule
\end{tabular}
}
\vspace{-0.3cm}
\end{table}

To assess the practical cost of \appname, we evaluate its runtime and storage overhead. 
Table~\ref{tab:cost_bound} reports the average results across the three evaluated models. 
\appname introduces a localization step that takes 23.81 seconds on average. 
Although this adds extra cost, RQ3 shows that localization improves editing effectiveness by identifying modules related to the package-validity boundary.
For editing, \appname requires 83.09 seconds on average, which is lower than ROME, MEMIT, and Full-FT. 
At inference time, \appname takes 11.30 seconds per case on average, which is comparable to ROME and lower than other baselines. 
\appname is also lightweight in storage, requiring only 731.03 KB of artifacts on average, approximately 20,000× smaller than full fine-tuning checkpoints. This is because \appname stores only LoRA updates on localized modules rather than large parameter edits or full checkpoints. 
Overall, \appname adds a small localization cost while achieving efficient editing, competitive inference time, and very small storage overhead.
These results show that \appname is practical for mitigating package hallucination.

\subsection{Impact of Editing on Model Performance}
\begin{table}[t]
\centering
\captionsetup{skip=2pt}
\caption{Code accuracy on HumanEval before and after editing.}
\label{tab:editing_drop}
\footnotesize
\setlength{\tabcolsep}{3pt}
\renewcommand{\arraystretch}{0.95}
\resizebox{\columnwidth}{!}{
\begin{tabular}{lccc|ccc}
\toprule
\multirow{2}{*}{Model}
& \multicolumn{3}{c|}{pass@1}
& \multicolumn{3}{c}{pass@10} \\
\cmidrule(lr){2-4} \cmidrule(lr){5-7}
& Orig. & Edited & $\Delta$
& Orig. & Edited & $\Delta$ \\
\midrule
DeepSeekCoder
& 68.96\% & 67.58\% & -1.38\%
& 82.32\% & 83.39\% & +1.07\% \\

Qwen3
& 23.29\% & 21.93\% & -1.36\%
& 51.83\% & 50.91\% & -0.92\% \\

Llama-3.1
& 41.52\% & 42.12\% & +0.60\%
& 69.51\% & 69.66\% & +0.15\% \\
\bottomrule
\end{tabular}
}
\vspace{-0.3cm}
\end{table}

While \appname effectively reduces package hallucination, it is important to examine whether editing affects the model's general code generation ability. 
To this end, we evaluate the original and edited models on HumanEval. 
Following the main evaluation setup, each model has four edited versions, obtained by editing on the four edit folds separately. 
We evaluate all four edited models on HumanEval and report the average pass@1 and pass@10 results. 
Table~\ref{tab:editing_drop} presents the results.

The results show that \appname remains stable on clean code generation inputs, with most changes within a $\pm$2\% margin. 
Overall, \appname mitigates package hallucination with minimal impact on the original capabilities of the model.

\subsection{Threats to Validity}
\noindent\textbf{Internal Validity:}
A potential internal threat comes from package extraction. 
We rely on regular expressions to extract package names from model outputs. 
Due to the inherent instability of LLM outputs, the extracted packages may not always follow the expected format. 
To reduce this threat, we constrain the output format in the prompt and use one-shot prompting to guide generation.
Another concern is the stability of evaluation. 
Since LLM generation involves randomness, the measured hallucination rates may exhibit slight variance. 
We address this by generating five responses for each prompt, and reporting average results over four edit folds.
The module localization step of \appname may also introduce uncertainty. 
Although \appname identifies modules related to package boundaries using a risk-aware localization objective, the results may still be affected by noisy gradients or the selected edit cases. 
To mitigate this threat, we aggregate localization signals over multiple edit cases. 

\noindent\textbf{External Validity:}
Our evaluation is conducted on three open-source LLMs from different model families. However, the results may not directly generalize to other LLMs. We focus on open-source models because they provide stable access after release, allowing historical versions to be revisited and reproduced. In contrast, commercial LLMs are continuously updated and may become unavailable over time.
Another external threat comes from the package ecosystem studied in this paper. 
We focus on Python packages from PyPI because PyPI provides public metadata for package validity checking. 
Future work can extend \appname to additional package ecosystems and programming languages (e.g., Java) to further evaluate its generality.
Furthermore, we evaluate \appname on three package-related tasks: package recommendation, code generation, and \texttt{pip install} recommendation. Although these tasks cover common package-related development scenarios, real-world dependency management also involves configuration files, build scripts, and repository-level workflows. Future work can extend \appname to these broader settings.

\section{Related Work}
\label{sec:relatedwork}
\subsection{Package Hallucination}
Prior work has studied hallucination in natural language generation and general LLM settings, such as dialogue, summarization, and factual question answering~\cite{ji2023survey,huang2025survey,zhang2025siren,alansari2026large}. 
Recent software engineering studies show that hallucination also appears in code generation, where models may produce incorrect APIs, non-existent libraries, or flawed implementation logic~\cite{liu2024exploring,tian2025codehalu,zhang2025llm}; documentation grounding has also been explored to mitigate API hallucinations~\cite{jain2025mitigating}. 
Package hallucination is a specific failure mode where an LLM recommends dependency names that do not exist or are unavailable in the target ecosystem. 
Empirical studies show that package hallucination is common across models, programming languages, model sizes, and task settings~\cite{spracklen2025we,krishna2025importing}, and HFuzzer further triggers such failures through phrase-based fuzzing~\cite{zhao2025hfuzzer}. 
This problem also raises software supply chain risks, since developers or coding agents may pass fabricated names to package managers, creating opportunities for package confusion, slop squatting, or malicious registration~\cite{ladisa2023sok,neupane2023beyond,kaplan2021survey,lanyado2024diving,spracklen2025we,krishna2025importing}.

Researchers have proposed several strategies to mitigate hallucinations~\cite{tonmoy2024comprehensive,alansari2026large}. 
Retrieval methods ground generation with external knowledge sources, self-correction prompts models to verify and refine their outputs, and model adaptation changes model behavior through fine-tuning or similar training procedures~\cite{jain2025mitigating,ji2023towards,madaan2023self,spracklen2025we}. 
Inference-time and post-hoc editing methods, such as TruthX and PURR, mitigate hallucinations by intervening in internal representations or revising generated outputs without full retraining~\cite{zhang2024truthx,chen2023purrefficientlyeditinglanguage}. 
However, these methods often depend on external resources, require repeated inference, revise outputs without correcting the model's default behavior, or incur high training costs. 
In contrast, we address package hallucination through localized model editing with a small set of editing examples.

\subsection{Model Editing}

Model editing, also known as knowledge editing, aims to correct specific knowledge or behaviors in pretrained models without full retraining~\cite{sinitsin2020editable,zhu2020modifying,wang2024detoxifying}. 
A successful edit should fix the target error while preserving the model's original capabilities, and is commonly evaluated by reliability, generalization, locality, and portability~\cite{zhang2024comprehensive,wang2024easyedit}. 
Existing methods can be broadly divided into two groups. 
Weight-preserving methods, such as SERAC, GRACE, and IKE, rely on external memories or in-context demonstrations to override selected predictions without modifying model parameters~\cite{mitchell2022memory,hartvigsen2023aging,zheng2023edit}. 
Weight-modified methods directly update model parameters through learned editors or localized changes. 
For example, KnowledgeEditor and MEND learn auxiliary networks to generate parameter updates~\cite{de2021editing,mitchell2022mend}, while later studies show that aggressive or sequential editing may harm general abilities or cause forgetting~\cite{gu2024modeleditingharms,gupta2024catastrophic}.

Among weight-modified methods, localized editing methods are especially relevant to our work.
Motivated by evidence that knowledge is encoded in specific transformer components~\cite{geva2021transformer,dai2022knowledge}, ROME, MEMIT, and PMET edit selected internal modules to rewrite factual associations at different scales~\cite{meng2022rome,meng2023memit,li2024pmet}. 
Recent work further extends model editing beyond factual triples, including detoxification, code LLM robustness, and editing for LLMs4Code~\cite{wang2024detoxifying,liu2025creme,li2025model}. 
However, package hallucination remains unexplored in model editing. 
Unlike factual updates that rewrite isolated subject-object associations, package hallucination requires correcting a task-conditioned package-validity boundary over package names. 
Our work targets this package-validity boundary editing problem.

\section{Conclusion}
\label{sec:conclusion}
In this paper, we presented \textbf{\appname}, a lightweight localized model editing framework for mitigating package hallucination in LLMs. 
\appname formulates package hallucination as a package-validity boundary editing problem, where the goal is to improve the model's distinction between valid and hallucinated package names. 
It first locates modules related to package hallucination through risk-aware localization, and then edits only these modules with lightweight LoRA adapters and a boundary-aware objective.
We evaluated \appname on three representative open-source LLMs: DeepSeekCoder, Qwen3, and Llama-3.1. 
The results show that \appname effectively reduces package hallucination in package recommendation, lowering Package-HR by 79.9\% on edit prompts and by 65.4\% on unseen prompts. 
The edited package-validity boundary also transfers to other package-related tasks, reducing Package-HR by 12.8\% in code generation and by 34.0\% in \texttt{pip install} recommendation. 
Compared with existing baselines, \appname better balances hallucination reduction and valid package preservation, while several baselines reduce hallucinations by suppressing package generation.
Ablation and cost analyses further show that each component contributes to \appname's effectiveness, and the framework is practical and lightweight. 
Overall, \appname provides a practical step toward mitigating package hallucinations in LLMs without full model retraining or external retrieval pipelines. 
Future work will extend \appname to more package ecosystems and broader dependency management scenarios.

\balance
\bibliographystyle{IEEEtran}
\bibliography{references}

@article{ji2023survey,
  title={Survey of hallucination in natural language generation},
  author={Ji, Ziwei and Lee, Nayeon and Frieske, Rita and Yu, Tiezheng and Su, Dan and Xu, Yan and Ishii, Etsuko and Bang, Ye Jin and Madotto, Andrea and Fung, Pascale},
  journal={ACM computing surveys},
  volume={55},
  number={12},
  pages={1--38},
  year={2023},
  publisher={ACM New York, NY}
}

@article{zhang2025siren,
  title={Siren’s Song in the AI Ocean: A Survey on Hallucination in Large Language Models},
  author={Zhang, Yue and Li, Yafu and Cui, Leyang and Cai, Deng and Liu, Lemao and Fu, Tingchen and Huang, Xinting and Zhao, Enbo and Zhang, Yu and Chen, Yulong and others},
  journal={Computational Linguistics},
  volume={51},
  number={4},
  pages={1373--1418},
  year={2025}
}

@article{liu2024exploring,
  title={Exploring and evaluating hallucinations in llm-powered code generation},
  author={Liu, Fang and Liu, Yang and Shi, Lin and Huang, Houkun and Wang, Ruifeng and Yang, Zhen and Zhang, Li and Li, Zhongqi and Ma, Yuchi},
  journal={arXiv e-prints},
  pages={arXiv--2404},
  year={2024}
}

@inproceedings{tian2025codehalu,
  title={Codehalu: Investigating code hallucinations in llms via execution-based verification},
  author={Tian, Yuchen and Yan, Weixiang and Yang, Qian and Zhao, Xuandong and Chen, Qian and Wang, Wen and Luo, Ziyang and Ma, Lei and Song, Dawn},
  booktitle={Proceedings of the AAAI Conference on Artificial Intelligence},
  volume={39},
  number={24},
  pages={25300--25308},
  year={2025}
}

@article{krishna2025importing,
  title={Importing phantoms: Measuring llm package hallucination vulnerabilities},
  author={Krishna, Arjun and Galinkin, Erick and Derczynski, Leon and Martin, Jeffrey},
  journal={arXiv preprint arXiv:2501.19012},
  year={2025}
}

@inproceedings{spracklen2025we,
  title={We have a package for you! a comprehensive analysis of package hallucinations by code generating {LLMs}},
  author={Spracklen, Joseph and Wijewickrama, Raveen and Sakib, AHM Nazmus and Maiti, Anindya and Viswanath, Bimal},
  booktitle={34th USENIX Security Symposium (USENIX Security 25)},
  pages={3687--3706},
  year={2025}
}

@inproceedings{jain2025mitigating,
  title={On mitigating code llm hallucinations with api documentation},
  author={Jain, Nihal and Kwiatkowski, Robert and Ray, Baishakhi and Ramanathan, Murali Krishna and Kumar, Varun},
  booktitle={2025 IEEE/ACM 47th International Conference on Software Engineering: Software Engineering in Practice (ICSE-SEIP)},
  pages={237--248},
  year={2025},
  organization={IEEE}
}

@article{zhao2025hfuzzer,
  title={Hfuzzer: Testing large language models for package hallucinations via phrase-based fuzzing},
  author={Zhao, Yukai and Wu, Menghan and Hu, Xing and Xia, Xin},
  journal={arXiv preprint arXiv:2509.23835},
  year={2025}
}

@article{huang2025survey,
  title={A survey on hallucination in large language models: Principles, taxonomy, challenges, and open questions},
  author={Huang, Lei and Yu, Weijiang and Ma, Weitao and Zhong, Weihong and Feng, Zhangyin and Wang, Haotian and Chen, Qianglong and Peng, Weihua and Feng, Xiaocheng and Qin, Bing and others},
  journal={ACM Transactions on Information Systems},
  volume={43},
  number={2},
  pages={1--55},
  year={2025},
  publisher={ACM New York, NY}
}

@article{alansari2026large,
  title={Large language models hallucination: A comprehensive survey},
  author={Alansari, Aisha and Luqman, Hamzah},
  journal={Computer Science Review},
  volume={61},
  pages={100970},
  year={2026},
  publisher={Elsevier}
}

@article{zhang2025llm,
  title={Llm hallucinations in practical code generation: Phenomena, mechanism, and mitigation},
  author={Zhang, Ziyao and Wang, Chong and Wang, Yanlin and Shi, Ensheng and Ma, Yuchi and Zhong, Wanjun and Chen, Jiachi and Mao, Mingzhi and Zheng, Zibin},
  journal={Proceedings of the ACM on Software Engineering},
  volume={2},
  number={ISSTA},
  pages={481--503},
  year={2025},
  publisher={ACM New York, NY, USA}
}

@misc{lanyado2024diving,
  author       = {Bar Lanyado},
  title        = {Diving Deeper into AI Package Hallucinations},
  year         = {2024},
  howpublished = {\url{https://www.lasso.security/blog/ai-package-hallucinations}},
}

@inproceedings{ladisa2023sok,
  title={Sok: Taxonomy of attacks on open-source software supply chains},
  author={Ladisa, Piergiorgio and Plate, Henrik and Martinez, Matias and Barais, Olivier},
  booktitle={2023 IEEE Symposium on Security and Privacy (SP)},
  pages={1509--1526},
  year={2023},
  organization={IEEE}
}

@inproceedings{neupane2023beyond,
  title={Beyond typosquatting: an in-depth look at package confusion},
  author={Neupane, Shradha and Holmes, Grant and Wyss, Elizabeth and Davidson, Drew and De Carli, Lorenzo},
  booktitle={32nd USENIX security symposium (USENIX security 23)},
  pages={3439--3456},
  year={2023}
}

@inproceedings{kaplan2021survey,
  title={A survey on common threats in npm and pypi registries},
  author={Kaplan, Berkay and Qian, Jingyu},
  booktitle={International Workshop on Deployable Machine Learning for Security Defense},
  pages={132--156},
  year={2021},
  organization={Springer}
}

@article{tonmoy2024comprehensive,
  title={A comprehensive survey of hallucination mitigation techniques in large language models},
  author={Tonmoy, SM and Zaman, SM and Jain, Vinija and Rani, Anku and Rawte, Vipula and Chadha, Aman and Das, Amitava},
  journal={arXiv preprint arXiv:2401.01313},
  year={2024}
}

@inproceedings{ji2023towards,
  title={Towards mitigating LLM hallucination via self reflection},
  author={Ji, Ziwei and Yu, Tiezheng and Xu, Yan and Lee, Nayeon and Ishii, Etsuko and Fung, Pascale},
  booktitle={Findings of the Association for Computational Linguistics: EMNLP 2023},
  pages={1827--1843},
  year={2023}
}

@article{madaan2023self,
  title={Self-refine: Iterative refinement with self-feedback},
  author={Madaan, Aman and Tandon, Niket and Gupta, Prakhar and Hallinan, Skyler and Gao, Luyu and Wiegreffe, Sarah and Alon, Uri and Dziri, Nouha and Prabhumoye, Shrimai and Yang, Yiming and others},
  journal={Advances in neural information processing systems},
  volume={36},
  pages={46534--46594},
  year={2023}
}

@article{yang2025qwen3,
  title={Qwen3 technical report},
  author={Yang, An and Li, Anfeng and Yang, Baosong and Zhang, Beichen and Hui, Binyuan and Zheng, Bo and Yu, Bowen and Gao, Chang and Huang, Chengen and Lv, Chenxu and others},
  journal={arXiv preprint arXiv:2505.09388},
  year={2025}
}

@article{grattafiori2024llama,
  title={The llama 3 herd of models},
  author={Grattafiori, Aaron and Dubey, Abhimanyu and Jauhri, Abhinav and Pandey, Abhinav and Kadian, Abhishek and Al-Dahle, Ahmad and Letman, Aiesha and Mathur, Akhil and Schelten, Alan and Vaughan, Alex and others},
  journal={arXiv preprint arXiv:2407.21783},
  year={2024}
}

@article{lewis2020retrieval,
  title={Retrieval-augmented generation for knowledge-intensive nlp tasks},
  author={Lewis, Patrick and Perez, Ethan and Piktus, Aleksandra and Petroni, Fabio and Karpukhin, Vladimir and Goyal, Naman and K{\"u}ttler, Heinrich and Lewis, Mike and Yih, Wen-tau and Rockt{\"a}schel, Tim and others},
  journal={Advances in neural information processing systems},
  volume={33},
  pages={9459--9474},
  year={2020}
}

@article{hu2022lora,
  title={Lora: Low-rank adaptation of large language models.},
  author={Hu, Edward J and Shen, Yelong and Wallis, Phillip and Allen-Zhu, Zeyuan and Li, Yuanzhi and Wang, Shean and Wang, Liang and Chen, Weizhu and others},
  journal={Iclr},
  volume={1},
  number={2},
  pages={3},
  year={2022}
}

@article{mitchell2022mend,
  title={Fast model editing at scale},
  author={Mitchell, Eric and Lin, Charles and Bosselut, Antoine and Finn, Chelsea and Manning, Christopher D},
  journal={arXiv preprint arXiv:2110.11309},
  year={2021}
}

@article{meng2022rome,
  title={Locating and editing factual associations in gpt},
  author={Meng, Kevin and Bau, David and Andonian, Alex and Belinkov, Yonatan},
  journal={Advances in neural information processing systems},
  volume={35},
  pages={17359--17372},
  year={2022}
}

@article{meng2023memit,
  title={Mass-editing memory in a transformer},
  author={Meng, Kevin and Sharma, Arnab Sen and Andonian, Alex and Belinkov, Yonatan and Bau, David},
  journal={arXiv preprint arXiv:2210.07229},
  year={2022}
}

@inproceedings{wang2024detoxifying,
  title={Detoxifying large language models via knowledge editing},
  author={Wang, Mengru and Zhang, Ningyu and Xu, Ziwen and Xi, Zekun and Deng, Shumin and Yao, Yunzhi and Zhang, Qishen and Yang, Linyi and Wang, Jindong and Chen, Huajun},
  booktitle={Proceedings of the 62nd Annual Meeting of the Association for Computational Linguistics (Volume 1: Long Papers)},
  pages={3093--3118},
  year={2024}
}

@article{pearce2025asleep,
  title={Asleep at the keyboard? assessing the security of github copilot’s code contributions},
  author={Pearce, Hammond and Ahmad, Baleegh and Tan, Benjamin and Dolan-Gavitt, Brendan and Karri, Ramesh},
  journal={Communications of the ACM},
  volume={68},
  number={2},
  pages={96--105},
  year={2025},
  publisher={ACM New York, NY, USA}
}

@inproceedings{perry2023users,
  title={Do users write more insecure code with ai assistants?},
  author={Perry, Neil and Srivastava, Megha and Kumar, Deepak and Boneh, Dan},
  booktitle={Proceedings of the 2023 ACM SIGSAC conference on computer and communications security},
  pages={2785--2799},
  year={2023}
}

@inproceedings{ohm2020backstabber,
  title={Backstabber’s knife collection: A review of open source software supply chain attacks},
  author={Ohm, Marc and Plate, Henrik and Sykosch, Arnold and Meier, Michael},
  booktitle={International Conference on Detection of Intrusions and Malware, and Vulnerability Assessment},
  pages={23--43},
  year={2020},
  organization={Springer}
}

@inproceedings{de2021editing,
  title={Editing factual knowledge in language models},
  author={De Cao, Nicola and Aziz, Wilker and Titov, Ivan},
  booktitle={Proceedings of the 2021 conference on empirical methods in natural language processing},
  pages={6491--6506},
  year={2021}
}

@article{sinitsin2020editable,
  title={Editable neural networks},
  author={Sinitsin, Anton and Plokhotnyuk, Vsevolod and Pyrkin, Dmitriy and Popov, Sergei and Babenko, Artem},
  journal={arXiv preprint arXiv:2004.00345},
  year={2020}
}

@article{zhu2020modifying,
  title={Modifying memories in transformer models},
  author={Zhu, Chen and Rawat, Ankit Singh and Zaheer, Manzil and Bhojanapalli, Srinadh and Li, Daliang and Yu, Felix and Kumar, Sanjiv},
  journal={arXiv preprint arXiv:2012.00363},
  year={2020}
}

@inproceedings{mitchell2022memory,
  title={Memory-based model editing at scale},
  author={Mitchell, Eric and Lin, Charles and Bosselut, Antoine and Manning, Christopher D and Finn, Chelsea},
  booktitle={International Conference on Machine Learning},
  pages={15817--15831},
  year={2022},
  organization={PMLR}
}

@inproceedings{geva2021transformer,
  title={Transformer feed-forward layers are key-value memories},
  author={Geva, Mor and Schuster, Roei and Berant, Jonathan and Levy, Omer},
  booktitle={Proceedings of the 2021 Conference on Empirical Methods in Natural Language Processing},
  pages={5484--5495},
  year={2021}
}

@inproceedings{dai2022knowledge,
  title={Knowledge neurons in pretrained transformers},
  author={Dai, Damai and Dong, Li and Hao, Yaru and Sui, Zhifang and Chang, Baobao and Wei, Furu},
  booktitle={Proceedings of the 60th Annual Meeting of the Association for Computational Linguistics (Volume 1: Long Papers)},
  pages={8493--8502},
  year={2022}
}

@article{zhang2024comprehensive,
  title={A comprehensive study of knowledge editing for large language models},
  author={Zhang, Ningyu and Yao, Yunzhi and Tian, Bozhong and Wang, Peng and Deng, Shumin and Wang, Mengru and Xi, Zekun and Mao, Shengyu and Zhang, Jintian and Ni, Yuansheng and others},
  journal={arXiv preprint arXiv:2401.01286},
  year={2024}
}

@inproceedings{wang2024easyedit,
  title={Easyedit: An easy-to-use knowledge editing framework for large language models},
  author={Wang, Peng and Zhang, Ningyu and Tian, Bozhong and Xi, Zekun and Yao, Yunzhi and Xu, Ziwen and Wang, Mengru and Mao, Shengyu and Wang, Xiaohan and Cheng, Siyuan and others},
  booktitle={Proceedings of the 62nd Annual Meeting of the Association for Computational Linguistics (Volume 3: System Demonstrations)},
  pages={82--93},
  year={2024}
}

@article{hartvigsen2023aging,
  title={Aging with grace: Lifelong model editing with discrete key-value adaptors},
  author={Hartvigsen, Tom and Sankaranarayanan, Swami and Palangi, Hamid and Kim, Yoon and Ghassemi, Marzyeh},
  journal={Advances in Neural Information Processing Systems},
  volume={36},
  pages={47934--47959},
  year={2023}
}

@inproceedings{zheng2023edit,
  title={Can we edit factual knowledge by in-context learning?},
  author={Zheng, Ce and Li, Lei and Dong, Qingxiu and Fan, Yuxuan and Wu, Zhiyong and Xu, Jingjing and Chang, Baobao},
  booktitle={Proceedings of the 2023 Conference on Empirical Methods in Natural Language Processing},
  pages={4862--4876},
  year={2023}
}

@inproceedings{gu2024modeleditingharms,
  title={Model editing harms general abilities of large language models: Regularization to the rescue},
  author={Gu, Jia-Chen and Xu, Hao-Xiang and Ma, Jun-Yu and Lu, Pan and Ling, Zhen-Hua and Chang, Kai-Wei and Peng, Nanyun},
  booktitle={Proceedings of the 2024 Conference on Empirical Methods in Natural Language Processing},
  pages={16801--16819},
  year={2024}
}

@inproceedings{gupta2024catastrophic,
  title={Model editing at scale leads to gradual and catastrophic forgetting},
  author={Gupta, Akshat and Rao, Anurag and Anumanchipalli, Gopala},
  booktitle={Findings of the Association for Computational Linguistics: ACL 2024},
  pages={15202--15232},
  year={2024}
}

@article{liu2025creme,
  title={CREME: Robustness Enhancement of Code LLMs via Layer-Aware Model Editing},
  author={Liu, Shuhan and Hu, Xing and Huang, Kerui and Yang, Xiaohu and Lo, David and Xia, Xin},
  journal={arXiv preprint arXiv:2507.16407},
  year={2025}
}

@inproceedings{wang2023review,
  title={A review on code generation with llms: Application and evaluation},
  author={Wang, Jianxun and Chen, Yixiang},
  booktitle={2023 IEEE International Conference on Medical Artificial Intelligence (MedAI)},
  pages={284--289},
  year={2023},
  organization={IEEE}
}

@inproceedings{jimenez2024swe,
  title={Swe-bench: Can language models resolve real-world github issues?},
  author={Jimenez, Carlos E and Yang, John and Wettig, Alexander and Yao, Shunyu and Pei, Kexin and Press, Ofir and Narasimhan, Karthik},
  booktitle={International Conference on Learning Representations},
  volume={2024},
  pages={54107--54157},
  year={2024}
}

@article{roziere2023code,
  title={Code llama: Open foundation models for code},
  author={Roziere, Baptiste and Gehring, Jonas and Gloeckle, Fabian and Sootla, Sten and Gat, Itai and Tan, Xiaoqing Ellen and Adi, Yossi and Liu, Jingyu and Sauvestre, Romain and Remez, Tal and others},
  journal={arXiv preprint arXiv:2308.12950},
  year={2023}
}

@inproceedings{sun2025source,
  title={Source code summarization in the era of large language models},
  author={Sun, Weisong and Miao, Yun and Li, Yuekang and Zhang, Hongyu and Fang, Chunrong and Liu, Yi and Deng, Gelei and Liu, Yang and Chen, Zhenyu},
  booktitle={2025 IEEE/ACM 47th International Conference on Software Engineering (ICSE)},
  pages={1882--1894},
  year={2025},
  organization={IEEE}
}

@article{alhanahnah2024depsrag,
  title={Depsrag: Towards agentic reasoning and planning for software dependency management},
  author={Alhanahnah, Mohannad and Boshmaf, Yazan},
  journal={arXiv preprint arXiv:2405.20455},
  year={2024}
}

@article{latendresse2025robust,
  title={How Robust are LLM-Generated Library Imports? An Empirical Study using Stack Overflow},
  author={Latendresse, Jasmine and Khatoonabadi, SayedHassan and Shihab, Emad},
  journal={arXiv preprint arXiv:2507.10818},
  year={2025}
}

@article{gao2025current,
  title={The current challenges of software engineering in the era of large language models},
  author={Gao, Cuiyun and Hu, Xing and Gao, Shan and Xia, Xin and Jin, Zhi},
  journal={ACM Transactions on Software Engineering and Methodology},
  volume={34},
  number={5},
  pages={1--30},
  year={2025},
  publisher={ACM New York, NY}
}

@misc{pypi,
  title        = {{PyPI}: The Python Package Index},
  author       = {{Python Software Foundation}},
  year         = {2026},
  howpublished = {\url{https://pypi.org/}},
  note         = {Accessed: 2026-06-14}
}

@inproceedings{li2025model,
  title={Model editing for llms4code: How far are we?},
  author={Li, Xiaopeng and Wang, Shangwen and Li, Shasha and Ma, Jun and Yu, Jie and Liu, Xiaodong and Wang, Jing and Ji, Bin and Zhang, Weimin},
  booktitle={2025 IEEE/ACM 47th International Conference on Software Engineering (ICSE)},
  pages={937--949},
  year={2025},
  organization={IEEE}
}

@inproceedings{li2024pmet,
  title={Pmet: Precise model editing in a transformer},
  author={Li, Xiaopeng and Li, Shasha and Song, Shezheng and Yang, Jing and Ma, Jun and Yu, Jie},
  booktitle={Proceedings of the AAAI Conference on Artificial Intelligence},
  volume={38},
  number={17},
  pages={18564--18572},
  year={2024}
}

@inproceedings{wu2023depn,
  title={Depn: Detecting and editing privacy neurons in pretrained language models},
  author={Wu, Xinwei and Li, Junzhuo and Xu, Minghui and Dong, Weilong and Wu, Shuangzhi and Bian, Chao and Xiong, Deyi},
  booktitle={Proceedings of the 2023 Conference on Empirical Methods in Natural Language Processing},
  pages={2875--2886},
  year={2023}
}

@article{guo2024deepseek,
  title={DeepSeek-Coder: when the large language model meets programming--the rise of code intelligence},
  author={Guo, Daya and Zhu, Qihao and Yang, Dejian and Xie, Zhenda and Dong, Kai and Zhang, Wentao and Chen, Guanting and Bi, Xiao and Wu, Yifan and Li, YK and others},
  journal={arXiv preprint arXiv:2401.14196},
  year={2024}
}

@inproceedings{huang2025can,
  title={Can knowledge editing really correct hallucinations?},
  author={Huang, Baixiang and Chen, Canyu and Xu, Xiongxiao and Payani, Ali and Shu, Kai},
  booktitle={International Conference on Learning Representations},
  volume={2025},
  pages={88116--88149},
  year={2025}
}

@inproceedings{zhang2024truthx,
  title={Truthx: Alleviating hallucinations by editing large language models in truthful space},
  author={Zhang, Shaolei and Yu, Tian and Feng, Yang},
  booktitle={Proceedings of the 62nd Annual Meeting of the Association for Computational Linguistics (Volume 1: Long Papers)},
  pages={8908--8949},
  year={2024}
}

@article{chen2023purrefficientlyeditinglanguage,
  title={Purr: Efficiently editing language model hallucinations by denoising language model corruptions},
  author={Chen, Anthony and Pasupat, Panupong and Singh, Sameer and Lee, Hongrae and Guu, Kelvin},
  journal={arXiv preprint arXiv:2305.14908},
  year={2023}
}

@article{welleck2019neural,
  title={Neural text generation with unlikelihood training},
  author={Welleck, Sean and Kulikov, Ilia and Roller, Stephen and Dinan, Emily and Cho, Kyunghyun and Weston, Jason},
  journal={arXiv preprint arXiv:1908.04319},
  year={2019}
}

\end{document}